\documentclass[preprint,showpacs,preprintnumbers,amsmath,amssymb,floatfix]{revtex4}
\usepackage{graphicx}% Include figure files
\usepackage{dcolumn}% Align table columns on decimal point
\usepackage{bm}% bold math
\usepackage{longtable}
\begin{document}

\title{ Trends in ferromagnetism, hole localization, and acceptor level 
depth for Mn substitution in
GaN, GaP, GaAs and GaSb}
\author{Priya Mahadevan and Alex Zunger \\
National Renewable Energy Laboratory, Golden 80401}
\date{\today}

\begin{abstract}
We examine the intrinsic mechanism of ferromagnetism in dilute magnetic semiconductors
by analyzing the trends in the electronic structure as the host is changed from GaN to GaP, GaAs and GaSb,
keeping the transition metal impurity fixed. In contrast with earlier interpretations which
depended on the host semiconductor, we found that a single mechanism is sufficient to 
explain the ferromagnetic stabilization energy for the entire series.
\end{abstract}


\maketitle

Mn substitution on the Ga site of GaX III-V semiconductors 
with X=N, P, As and Sb
creates a hole-producing acceptor level E(0/-) and renders the
system ferromagnetic. Mean field 
models \cite{ohno, dietl,others}
aim to explain this effect by postulating that the Mn induced hole is a
delocalized, host-like shallow acceptor (analogous to that 
introduced by Zn doping in GaAs) which couples to the 
localized $S$=5/2 magnetic moment of Mn. Since, however, it is now known that
Mn in GaN introduces a {\it deep} ($\sim$ midgap) acceptor \cite{graf},
with highly {\it localized} hole states \cite{mark,prblong}, a different mechanism needs to be
invoked within previous approaches to explain ferromagnetism in nitrides \cite{dietl} as opposed to arsenides.
Using first-principles total-energy calculations of Mn in GaN, GaP, GaAs and
GaSb, we show that, in fact, a single mechanism suffices to reveal 
systematic trends in the ferromagnetic stabilization energies in this entire series. The 
calculations reveal clear trends in (i) hole localization, (ii) acceptor level
depth and (iii) ferromagnetic stabilization energy, all increasing along the GaSb $\rightarrow$ 
GaAs $\rightarrow$ GaP $\rightarrow$ GaN series. These trends reflect an
enhanced coupling between $p$-$d$ hybrids on different Mn sites along the
series. 

Experimentally, the trends along the GaX series are often clouded by
growth-induced imperfections, including the formation of 
clusters \cite{clusters}, precipitates \cite{precip1,precip2}, 
presence of anti-site defects \cite{asite} and
non-substitutional impurity sites \cite{interst}. To
clarify the {\it underlying intrinsic mechanism} of ferromagnetism (FM),
we intentionally simplify the problem drastically, considering defect-free cases
with only substitutional impurities. This approach reveals clear trends.
In order to address the issues of trends, we have constructed 
64 atom supercells of zinc-blende (ZB) GaN, GaP, GaAs and GaSb
in which we replace one or two Ga atoms by Mn. 
The lattice constant of the Mn-containing supercells are taken from 
Ref.~\cite{prblong}.
We have performed plane-wave pseudopotential total energy 
calculations \cite{ihm_zunger}  using the GGA exchange 
functional \cite {pw91} as implemented in the VASP \cite{vasp} code. We 
relaxed all atomic positions, calculating  the total energy 
for the neutral and negatively charged state of Mn$_{Ga}$ 
thus obtaining acceptor energies \cite{prblong} E(0/-).
In addition we calculated the total energy difference for two Mn spins aligned
ferromagnetically and antiferromagnetically, determining the energy difference, E$_{FM}$-E$_{AFM}$ - a measure of the  stability
of the ferromagnetic state.

Figure. 1 shows the calculated Mn $d$ projected local density of states for
neutral substitutional Mn in four ZB GaX compounds. The states are designated by 
$t_2$ or $e$ to denote the symmetry of the representations that they correspond
to, and by $+$ and $-$, to denote spin-up or spin-down respectively. The height of the
peaks reflect the degree of localization of the state within a
sphere of radius 1.2 $\AA$ about the Mn site. We find a {\it pair} of $t_2$ 
states in each spin channel, reflecting bonding and antibonding states 
formed by the $p$-$d$ interaction of Mn $d$ states with host 
derived anion $p$ states. The Fermi energy in all cases lies within the 
antibonding $t_+$ level, which in the neutral charge state of the 
impurity is partially occupied with two electrons (and therefore one hole).
We see that while in GaN:Mn, the hole-carrying $t_+^2$ level is strongly 
Mn-localized, the degree of Mn-localization of the hole level 
decreases along the series GaN $\rightarrow$ GaP $\rightarrow$ 
GaAs $\rightarrow$ GaSb. This is seen also in Fig. 2 which depicts the
wavefunction square of the antibonding $t_+$ state at $E_F$, evaluated at
$k$=0. In contrast to what is expected from host-like-hole models \cite{ohno,dietl,others}, 
we find significant weight of the hole only on a unit comprising Mn and its
nearest neighbors. This $t_+$ state has earlier been identified \cite{prblong} 
as the antibonding state arising
from $p$-$d$ interactions between the Mn $t_2$(d) states 
with the corresponding $t_2$(p) states localized on the anion nearest 
neighbors. (The antibonding nature is evident from the pocket of zero
charge density along the Mn-X bond). Concomitant with the reduced Mn 
localization on the  {\it antibonding} $t_+$ state at $E_F$, the {\it bonding} state
at 3-4 eV below the valence band maximum
exhibits (in Fig. 1) increased Mn localization along the same 
GaN $\rightarrow$ GaP $\rightarrow$ GaAs $\rightarrow$ GaSb series
( as a result of "$pd$~ level anticrossing" discussed in \cite{prblong}).

By adding an electron to the antibonding $t_+^2$ level, we convert 
Mn$^{3+}$($t^2$) to Mn$^{2+}$($t^3$), thus creating an acceptor transition. The corresponding
change E(0/-) in the total energy 
is depicted in Fig. 3
with respect to the calculated band edges \cite{boffset} of the unstrained 
zincblende solids. We see that the acceptor level is very deep in
GaN:Mn in agreement with experiment \cite{graf} and it becomes
progressively shallower as the anion X becomes heavier, also in agreement
with experiment \cite{schneider}.
Thus, moving along the series GaN $\rightarrow$ GaP
$\rightarrow$ GaAs $\rightarrow$ GaSb, the {\it antibonding}, hole-carrying
$t_+$ orbital becomes less Mn-localized, the acceptor level becomes
shallower, and the bonding orbitals located inside the valence band 
become more localized on Mn, as observed in photoemission experiments
\cite{pes}. The strong $d$ character of the hole in Mn doped GaX semiconductors has 
implications on the usefulness of these materials as a source of spin-polarized 
carriers (ferromagnetic injectors in spintronic devices). 
For instance  Mn in GaN has a deep acceptor level, which implies that it will not provide an easy
source of holes to make GaN $p$-type.

To assess the ferromagnetic stability, we compute the energy difference
$E_{FM}$ - $E_{AFM}$ between a ferromagnetic (FM) and an
antiferromagnetic (AFM) spin arrangement on the two Mn atoms in the
supercell. The two Mn atoms are positioned at first, second, third
and fourth fcc-neighbor positions ($n$=1,2,3,4 respectively); the
corresponding Mn-Mn exchange interaction strengths versus $n$ are depicted
in Fig. 4. The trends in our calculations are consistent with what is 
found earlier \cite{Sato}. Although the results  shown in Fig.~4 suggest a 
more rapid decay of the exchange interaction strengths for GaN compared to GaAs,
one finds that at fourth neighbor separation the interaction strengths are
comparable in agreement with what is found in Ref.~\cite{kang}.
However, the structure 
more relevant to experiment is the wurtzite structure in the case of GaN. 
Recent theoretical calculations suggest \cite{kang} 
that the exchange interaction  energies for Mn in wurtzite GaN fall off more
rapidly with distance than in the zincblende structure.
We see

1. Strong FM stabilization in GaN:Mn, despite the fact that the hole 
orbital is a highly localized (Figs. 1,2), deep acceptor (Fig. 3) state.
Even though our conclusion that GaN:Mn shows large ferromagnetic stabilization energies 
for some pairs of Mn atoms
parallels that of Dietl and Ohno \cite{dietl}, the mechanism behind it is
entirely different: Their model assumes a host-like
delocalized hole for all materials, on the basis of which they attributed trends in $J$ to 
{\it volume scaling}, $J$ $\sim$ $V^{-1}$=$R^{-3}$, leading to a large $J$
for the shortest bond-length material (here, GaN). We allow different materials
to exhibit different localizations, finding large FM stabilization energies in GaN {\it despite}
it not having host-like-hole states. Consistent with the deeper acceptor
level for GaN:Mn (Fig.~3) and its more localized hole-carrying orbital (Fig.~2),
it's spin-spin interaction is shorter-ranged (viz. 2nd and 3rd neighbors
in Fig.~4).

2. The exchange interaction strengths are large 
even for fourth neighbor Mn pair separations (Fig. 4). Thus,
{\it indirect} exchange (via the intervening anions) is at play.

3. Certain crystallographic orientations of the Mn-Mn pair are seen to have the
largest stabilization energies for the ferromagnetic state, e.g. the $<$110$>$ orientation akin to $n$=1 and 
$n$=4 neighbors, whereas $<$001$>$-oriented pairs ($n$=2) have the weakest
FM. This reflects the orientational dependence of the coupling matrix
elements between the two $t_2$ (pd) hybrids orbitals located on different Mn sites. 
The matrix elements and therefore the 
bonding is maximized when the $t_2$ (pd) hybrid orbitals point towards each other.
On the other hand in zincblende symmetry $e$-orbitals point $in-between$
the nearest neighbors, leading to vanishing matrix elements when the Mn atoms
occupy $<$110$>$ oriented lattice positions.
Indeed there can be several exchange paths, between the $t_2$ orbitals 
on the two Mn atoms, the most obvious choice being one mediated via the
host anion states {\it via} $p$-$d$ coupling, and the other mediated by 
$d$-$d$ interaction.  Our calculations help us distinguish which is the 
relevant exchange path.
For cases where the hole is in a level with $e$ symmetry,
such as GaAs:Fe$-$ \cite{prblong}, the ferromagnetic contribution from
the hole is found to be small. As the coupling to the host is absent for a hole
with $e$ symmetry, this suggests that
the relevant exchange path in these systems
is {\it via} $p$-$d$ coupling mediated by the host 
semiconductor states.

4. We conclude that the $p$-$d$ interaction couples the $d$ levels on Mn ion to the $p$-like 
dangling bond states of the Ga vacancy, thus creating $p$-$d$ hybrids localized on Mn and its
neighbors. The interaction of such partially occupied $t_+^2$ orbitals between different 
Mn sites stabilizes FM. The chemical trends in localization (Fig.2), acceptor energies (Fig. 3) and the
exchange interaction strengths (Fig. 4) reflect the position of the 
Mn $d$  orbital energies relative to the 
host band edges.

We acknowledge support from the Office of Naval Research.

%\bibliography{apl-final}

\begin{thebibliography}{}

\bibitem{ohno} 
T.~Dietl, H. Ohno, F. Matsukura, J. Cibert and D. Ferrand, Science {\bf 287}, 1019 (2000).

\bibitem{dietl} 
T. Dietl, Semicon. Sci. and Tech. {\bf 17}, 377 (2002).

\bibitem{others}
T. Dietl, F. Matsukura, and H. Ohno, Phys. Rev. B {\bf 66}, 033203 (2002);
J. K$\ddot{o}$nig, J. Schiemann, T. Jungwirth and A.H. MacDonald,
in {\it Electronic Structure and Magnetism of Complex Materials},
edited by D.J. Singh and D.A. Papaconstantopoulos, Springer verlag (2002).

\bibitem{graf} 
T. Graf, M. Gjukic, M.S. Brandt, M. Stuzmann, and O. Ambacher,
Appl. Phys. Lett. {\bf 81}, 5159 (2002).

\bibitem{mark} 
M. van Schilfgaarde and O.N. Mryasov, Phys. Rev. B {\bf 63}, 233205 (2001).

\bibitem{prblong}
P. Mahadevan and A. Zunger, Phys. Rev. B {\bf 69}, 115211 (2004).

\bibitem{clusters}
S. S. A. Seo, M. W. Kim, Y. S. Lee, T. W. Noh, Y. D. Park, G. T. Thaler,
M. E. Overberg, C. R. Abernathy and S.J. Pearton, Appl. Phys. Lett.
{\bf 82}, 4749 (2003).

\bibitem{precip1}
S. Dhar, O. Brandt, A. Trampert, L. Daweritz, K.J. Friedland, K.H. Ploog,
J. Keller, B. Beschoten, G. Guntherodt, Appl. Phys. Lett. {\bf 82}, 2077
(2003).

\bibitem{precip2}
F. Matsukura, E. Abe, Y. Ohno and H. Ohno, Appl. Surf. Sc. {\bf 159-160}, 
265 (2000).

\bibitem{asite} 
B. Grandidier, J.P. Nys, C. Delerue, D. Stievenard, Y. Higo and M. Tanaka,
Appl. Phys. Lett. {\bf 77}, 4001 (2000).

\bibitem{interst}
K.M. Yu, W. Walukiewicz, T. Wojtowicz, I. Kuryliszyn, X. Liu, Y. Sasaki,
and J.K. Furdyna, Phys. Rev. B {\bf 65}, 201303 (2002).

\bibitem{ihm_zunger}
J.~Ihm, A.~Zunger and M.L.~Cohen, J. Phys. C:{\bf 12}, 4409 (1979).

\bibitem{pw91}
J.P. Perdew and W. Wang, Phys. Rev. B {\bf 45}, 13244 (1992).

\bibitem{vasp}
G.~Kresse and J.~Furthm$\ddot{u}$ller, Phys. Rev. B. {\bf 54}, 11169 (1996);
G.~Kresse and J.~Furthm$\ddot{u}$ller, Comput. Mat. Sci. {\bf 6}, 15 (1996).

\bibitem{boffset}
S.H.~Wei and A.~Zunger, Appl. Phys. Lett. {\bf 72}, 2011 (1998).

\bibitem{schneider} 
J.~Schneider in {\it Defects in Semiconductors II, Symposium Proceedings}, Edt. S. Mahajan and J.W. Corbett, 
225 (North-Holland, 1983);
B.~Clerjaud, J. Phys. C {\bf 18}, 3615 (1985).

\bibitem{pes}
J.~Okabayashi, A. Kimura, O. Rader, T. Mizokawa, A. Fujimori,
T. Hayashi and M. Tanaka, Phys. Rev. B {\bf 58}, 4211 (1999).

\bibitem{Sato} 
K. Sato, P.H. Dederics and H. Katayama-Yoshida,
Europhys. Lett. {\bf 61}, 403 (2003).

\bibitem{kang}
J. Kang, K.C. Chang and H. Katayama-Yoshida, 
J. Superconductivity (in press).

\end{thebibliography}

\clearpage
\newpage

\begin{figure}
\caption{ Mn $d$ projected partial density of states for a single Mn in ZB GaN, GaP, GaAs and GaSb, where
the symmetry ($t_2$ and $e$ ) as well as the spin (+ and -) have been indicated. The shaded region represents
the $t_2^+$ states.
}
\end{figure}

\begin{figure}
\caption{Wavefunction squared plots of the hole wavefunction taken at $\Gamma$ point for Mn in 
the $<$110$>$ plane of ZB GaN, GaP, GaAs and
GaSb. The lowest contour (blue) represents 0.015 e/A$^3$ and the maximum contour (red) have been chosen so that the integrated
charge density is 90$\%$. Mn (Ga) atoms have been indicated by filled (open) black circles, while anion atoms are represented
by filled red circles.
}
\end{figure}

\begin{figure}
\caption{Calculated acceptor levels Mn(0/-) in III-V's. The host band edges are aligned according to the
calculated unstrained band offsets \cite{boffset}.
}
\end{figure}

\begin{figure}
\caption{Calculated exchange interaction strengths between Mn atoms
occupying 1st, 2nd, 3rd and 4th fcc-neighbor positions in a zincblende lattice of GaN, GaP, GaAs and
GaSb respectively.
}
\end{figure}

\end{document}